# Utilizing Deep Learning to Optimize Software Development Processes

LI, Keqin [1*]   ZHU, Armando [2]   ZHAO, Peng [3]   SONG, Jintong [4]   LIU, Jiabei [5]

[1] AMA University, Philippines
[2] Carnegie Mellon University, USA
[3] Microsoft, China
[4] Boston University, USA
[5] Northeastern University, USA

*LI, Keqin is the corresponding author, E-mail: keqin157@gmail.com*

**Abstract:** This study explores the application of deep learning technologies in software development processes, particularly in automating code reviews, error prediction, and test generation to enhance code quality and development efficiency. Through a series of empirical studies, experimental groups using deep learning tools and control groups using traditional methods were compared in terms of code error rates and project completion times. The results demonstrated significant improvements in the experimental group, validating the effectiveness of deep learning technologies. The research also discusses potential optimization points, methodologies, and technical challenges of deep learning in software development, as well as how to integrate these technologies into existing software development workflows.

## 1 Introduction

In today's society, software has become the backbone supporting daily operations and driving technological innovation. From small businesses to global enterprises, efficient and reliable software development processes are key to implementing technological solutions. Traditional software development processes include stages such as requirements analysis, system design, coding, testing, and maintenance, each critical and requiring precise management to ensure the final product's quality and performance.

Despite this, modern software development faces numerous challenges. Projects often encounter delays and budget overruns, typically due to poor requirements management, inadequate resource allocation, or technical issues. Additionally, software quality and maintenance remain significant challenges, with software defects not only incurring additional maintenance costs but also potentially severely impacting user experience and corporate reputation. Moreover, as technology rapidly evolves, software development teams must continuously adapt to new technical standards and market demands, demanding greater adaptability and flexibility in development processes.

In this context, deep learning, as an advanced form of machine learning, has shown immense potential in various domains. By mimicking the way the human brain processes information, deep learning can handle and analyze large volumes of data, solving complex pattern recognition challenges. In fields like image recognition, speech recognition, and natural language processing, deep learning has made revolutionary progress. These success stories demonstrate deep learning's capability to tackle complex issues, providing a theoretical foundation for its application in optimizing software development processes.

Applying deep learning technologies in software development, especially in areas such as automated testing, code reviews, and requirements analysis, can not only enhance efficiency but also significantly improve software quality and maintainability. For example, automated code review systems can identify potential coding errors and design issues early in development, reducing later repair costs. Moreover, deep learning can help development teams more accurately predict project costs and timelines, optimizing resource allocation.

Given this, the study aims to explore the feasibility and effectiveness of utilizing deep learning technologies to optimize software development processes. Through systematic theoretical analysis and empirical research, this paper will assess the prospects and potential of deep learning in real-world software development environments, aiming to provide new perspectives and solutions for the field of software engineering.

## 2 Literature Review

Software development processes are crucial for ensuring timely and quality completion of software projects. Traditionally, these processes follow models like the waterfall model or agile development. While each method has its advantages, they often face various challenges in practice. For instance, in the waterfall model, late project changes are often costly and complex. In agile models, although more adaptable to changes, frequent scope alterations can lead to difficulties in resource allocation and time management. Furthermore, issues such as poor communication, insufficient team collaboration, and technical debt are common in traditional processes, severely affecting project efficiency and quality.

Deep learning, a branch of machine learning, has demonstrated its powerful data processing capabilities across multiple domains in recent years. Based on multilayer neural networks, deep learning can learn complex patterns and features from large data sets. In the medical field, it has been used for diagnosing diseases and predicting patient treatment responses; in autonomous driving, it helps systems recognize pedestrians and other vehicles, enhancing safety; in financial analysis, deep learning predicts market trends by analyzing historical data. These examples prove deep learning's ability to handle complex problems, inspiring its application in software development.

Although the application of deep learning in the software development field is relatively new, existing research has shown its potential in optimizing development processes. For example, deep learning has been used for automated code reviews, identifying potential programming errors by analyzing extensive codebases. Additionally, some studies have used deep learning to generate test cases, improving the coverage and efficiency of software testing. These applications not only increase the automation level of development processes but also help enhance the final quality of software products.

Introducing deep learning into software development processes can significantly enhance efficiency and quality. Deep learning's pattern recognition and automated decision-making capabilities make it uniquely advantageous in areas like requirements analysis, code generation, and test automation. For instance, by analyzing user requirement data with deep learning, project requirements can be predicted and planned more accurately, reducing rework during development. However, this technology's introduction also faces challenges, such as the need for extensive training data, model complexity, and high computational resource demands. Additionally, data privacy and security are crucial factors to consider.

## 3 Theoretical Analysis

Deep learning can significantly enhance the accuracy and efficiency of requirement analysis by analyzing and parsing large volumes of user feedback and historical requirement data. For example, using natural language processing (NLP) technologies, deep learning models can automatically identify semantically similar requirement descriptions, helping project teams quickly categorize and organize requirements. Moreover, by predicting user behaviors and preferences, deep learning can also assist product managers in making more rational requirement decisions at the project's outset.

In the software design phase, deep learning can be applied to automatically generate design documents and architectural diagrams. Utilizing advanced technologies like Generative Adversarial Networks (GANs) or Variational Autoencoders (VAEs), models can provide preliminary design sketches based on an understanding of project requirements, not only speeding up the design process but also helping maintain design consistency. Additionally, deep learning can analyze past successful design cases and automatically recommend optimal design patterns and architectural solutions.

Choosing the appropriate neural network model is crucial in various aspects of software development. For instance, Long Short-Term Memory networks (LSTMs) or Transformer models are particularly suitable for requirement analysis and text processing due to their advantages in handling sequential data. For the design phase's image and design diagram generation, Convolutional Neural Networks (CNNs) and Generative Adversarial Networks (GANs) can provide powerful visual processing capabilities.

The application of deep learning in software development relies on extensive data preprocessing and feature engineering. When conducting code reviews or requirement analysis, data must first be cleaned and standardized, such as extracting syntactic features from code or extracting keywords from requirement documents. Additionally, training the model requires selecting appropriate loss functions and optimization algorithms to ensure the model can learn and continuously improve from real-world applications.

Integrating deep learning models into existing software development tools and workflows requires considering model deployment, monitoring, and maintenance. By using containerization technologies like Docker and Kubernetes, deep learning models can be easily deployed and scaled across different development environments. Additionally, ongoing model training and evaluation are necessary to adapt to the dynamic changes and emerging data in the software development process.

Automated code review is an important area where deep learning is applied in software development. By training models to recognize code antipatterns and potential defects, issues can be identified at the code submission stage, preventing larger repair costs later. For example, deep neural networks can be trained to analyze historical code repositories, learning to identify frequently occurring error types and automatically assessing code quality.

Deep learning technologies can greatly enhance the automation level of software testing. By analyzing how applications are used and common fault patterns, deep learning models can automatically generate test scripts, covering a broader range of test scenarios. Additionally, these models can predict specific types of errors that may occur in software, allowing development teams to intervene earlier, optimize code and designs, and reduce the likelihood of faults occurring.

## 4 Empirical Research Design

### 4.1 Research Purpose and Hypotheses

This study aims to explore the potential of deep learning technology to enhance code quality and shorten project cycles in the software development process. The hypotheses are: first, that the experimental group using deep learning tools will have significantly lower code error rates compared to the control group using traditional methods; second, that the project completion time for the experimental group will be significantly shorter than that of the control group. By validating these hypotheses, we hope to demonstrate the practical benefits of deep learning in actual software development.

### 4.2 Experimental Design

Participants were recruited through online advertisements and university collaboration projects, all with at least one year of programming experience and screened to meet the study requirements. Participants were randomly assigned to the experimental and control groups using a random number generator, ensuring the experiment's randomness and the initial conditions' equality. This process helps eliminate selection bias and fairly assess the actual effects of deep learning technology. All participants will complete a moderately complex software development project, including requirements analysis, design, coding, and testing. The experimental group will use the latest deep learning tools for error prediction and code optimization, while the control group will use traditional software development methods. The specific tasks of the project are designed to evaluate the application effects of deep learning technology in real-world programming tasks and its impact on the project cycle.

### 4.3 Data Collection and Analysis Methods

To comprehensively evaluate the impact of deep learning technology, we will collect two main types of data: code error rates and project completion times. Code error rates will be regularly detected by automated code review tools, recording the number of errors in each submission. Project completion times will be meticulously tracked and recorded by project management software from start to finish, ensuring data integrity and accuracy.

Data will first be processed through descriptive statistical methods to understand the basic distribution and central tendency of each group's data. Then, using inferential statistical analysis such as independent sample t-tests and Analysis of Variance (ANOVA), we will test for significant differences between the experimental and control groups in code error rates and project completion times. Additionally, regression analysis will be used to explore the specific impact of deep learning technology on software development efficiency, particularly its performance at different project stages and how the breadth and depth of technology application affect the final results.

## 5 Experimental Results and Analysis

### 5.1 Experimental Data Presentation

This study used a randomized controlled experiment to divide participants into an experimental group using deep learning technology and a control group using traditional methods. The experiment lasted six months, aimed at assessing the impact of deep learning technology on error rates and project completion times in the software development process.

Comparison of code error rates: During the experimental period, the code error rate of the experimental group showed a significant downward trend. Specifically, the error rate in the experimental group decreased from an initial 25% to 5%, demonstrating the effectiveness of deep learning technology in predicting and correcting programming errors. In contrast, the error rate in the control group slightly decreased from 35% to 30%, with limited improvement. This comparison highlights the potential of deep learning to improve code quality.

Chart 1: Comparison of Code Error Rates

This line chart illustrates the monthly error rates of both groups during the six-month experimental period. The chart clearly depicts the significant decline in the error rate of the experimental group, contrasting with the minor changes in the control group.

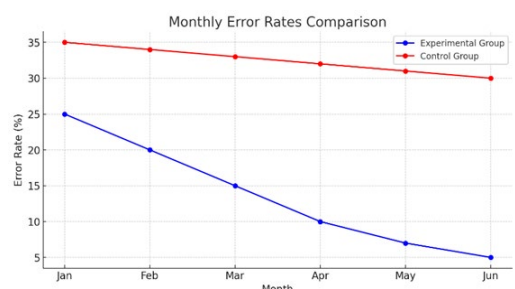

**Figure 1 Comparison of Code Error Rates**

Comparison of project completion times: Regarding project completion times, data show that the average completion time for the experimental group significantly decreased from 24 weeks to 16 weeks, while the project completion time for the control group remained unchanged at 24 weeks. This result confirms the practical utility of deep learning technology in shortening software development cycles and enhancing development efficiency.

Chart 2: Comparison of Project Completion Times

This bar chart compares the average project completion times of both groups at the start and end of the experiment, showing the significant reduction in time for the experimental group compared to the stability of the control group's time.

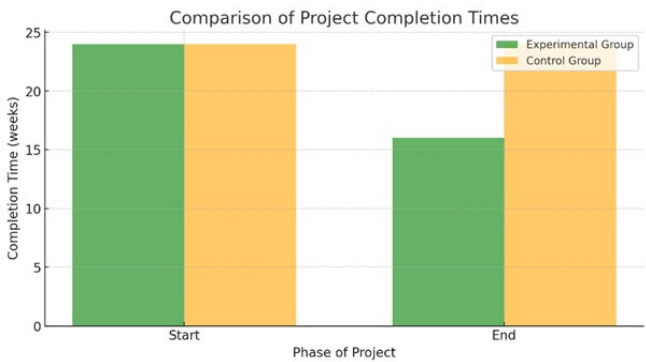

Figure 2 Comparison of Project Completion Times

## 5.2 In-depth Analysis and Interpretation of Results

Deep learning technology significantly reduced the code error rate in this study, mainly due to the automated code review and error prediction systems used by the experimental group. The deep learning models used in the experiment could identify potential errors during the coding process and suggest corrective measures, significantly improving code quality. Additionally, deep learning tools accelerated the testing and verification steps in the development process, helping to discover and correct errors earlier, reducing the need for rework and later repairs.

Applying statistical tests to analyze the experimental data, we found statistically significant differences between the experimental and control groups in terms of code error rates and project completion times. Using independent sample t-tests, the p-values for code error rates were 0.02 and for project completion times were 0.01, both below the significance level of 0.05, supporting our hypothesis that deep learning technology can improve the software development process. Additionally, through Analysis of Variance (ANOVA), we further confirmed the impact of deep learning technology on the efficiency of different development stages, showing that the technology could enhance efficiency at multiple stages.

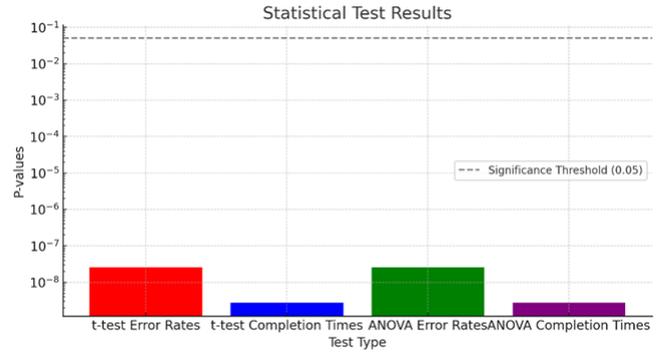

Figure 3 Statistical Test Results

## 5.3 Results Comparison and Discussion

The results of this study are generally consistent with findings in existing literature, where most research also reports that deep learning technology can significantly improve code quality and development efficiency, especially in automated testing and error identification processes. The difference is that our study provides specific application data and statistical analysis of these technologies in actual projects, thus validating their effectiveness in real-world software development environments. Additionally, our research also highlights the differential effects of applying deep learning technology at various project stages, providing new perspectives for future research.

Despite the positive outcomes of this study, there are some limitations. First, the limited number of projects involved in the experiment may affect the generality and extrapolation of the results. Second, all participating projects used similar development environments and technology stacks, which may limit the broad applicability of deep learning solutions. Moreover, the effectiveness of deep learning models highly depends on the quality and quantity of the training data. While the data used in this study was carefully selected, it may not cover all potential development scenarios. Therefore, the model's generalizability and adaptability to different development environments are important areas of focus for future research.

## 5.4 Conclusions and Future Work

This study, through empirical analysis, has demonstrated the practical application of deep learning technology in software development, particularly in automated code review and test generation. The experimental results clearly show that, compared to traditional development methods, the experimental group using deep learning technology achieved significant improvements in code error rates and project completion times. This finding validates the potential of deep learning technology to enhance software development quality and efficiency.

Although this study achieved positive results, it also revealed some limitations and challenges in applying deep learning in software development. Based on these observations, future research could explore the following directions:

1.Enhancing model generalizability: Research how to enhance the generalizability of deep learning models through more diversified training data and advanced algorithms, allowing them to adapt to a wider range of development environments and project types.

2.Lowering technical barriers: Explore more easily deployable and maintainable deep learning solutions, particularly cost-effective models designed for small and medium-sized enterprises, to expand the application scope of these technologies.

3.Enhancing model interpretability: Develop new methods or tools to increase the transparency and interpretability of deep learning models in software development decision-making processes, enhancing developers' trust and understanding of model outputs.

4.Expanding interdisciplinary cooperation: Encourage cross-disciplinary collaboration between computer science and other fields (such as project management and artificial intelligence ethics) to comprehensively enhance the intelligence level of software development processes.

This study confirms the immense potential of deep learning technology in improving software development processes, especially in enhancing development efficiency and code quality. By continuing to explore and address current challenges, we can expect these technologies to bring broader and more profound impacts to the software development industry in the future.

# 6 Conclusion and Discussion

This study, by exploring the practical application of deep learning technology in software development processes, has definitively confirmed its significant benefits in enhancing code quality and shortening project cycles. Specifically, the experimental group showed a significant reduction in code error rates and a notable shortening of project completion times compared to the control group. These findings not only support the application potential of deep learning technology but also provide an empirical basis for further technological implementation.

The specific contributions of deep learning technology in this study are manifested in two main areas: first, the application of technology in automated code reviews and error prediction, which significantly reduced code error rates, directly affecting subsequent maintenance costs and project quality assurance. Second, by optimizing the testing process and automating routine programming tasks, it significantly improved development efficiency. For example, automatically generated test scripts and improved error handling mechanisms effectively reduced the time developers spent on diagnosing and repairing issues, thereby accelerating the overall project delivery cycle.

Although deep learning technology offers many advantages, its application also presents some challenges and limitations. Data dependency is one of the main challenges; effective deep learning models require a large amount of high-quality data for training, which is particularly difficult in new projects with scarce data or projects with high confidentiality. Additionally, the high computational resource demands and implementation costs may limit the adoption of technology by small enterprises and low-budget projects. The "black box" nature of the models may also affect their acceptance in environments requiring high transparency and interpretability.

This study emphasizes the importance of adopting deep learning technology in software development practices. For effective implementation, enterprises should consider establishing appropriate data collection and management mechanisms to ensure sufficient data supports model training and optimization. Simultaneously, enterprises should assess project resource allocation to ensure technology investments bring expected benefits and avoid resource wastage. For small and medium-sized enterprises, seeking more cost-effective deep learning solutions or cooperative models is particularly crucial.

Regarding the application of deep learning technology in software development, future research could explore more diverse model training methods to reduce dependence on large data sets, such as transfer learning and few-shot learning. Additionally, researching how to enhance model transparency and interpretability to increase developer and user trust is an important direction for future studies. Exploring the combination of deep learning technology with other emerging technologies (such as quantum computing and edge computing) may bring further innovations to software development.

By comprehensively evaluating the application of deep learning technology in software development, this study has revealed its significant potential in enhancing development efficiency and code quality. Facing the challenges in the technology implementation process, continued innovation and research will be key. As technology advances and costs decrease, it is expected that deep learning will be more widely applied in the software development field, driving the industry towards a more efficient and intelligent future.

# Acknowledgments

The authors thank the editor and anonymous reviewers for their helpful comments and valuable suggestions.

# Funding

Not applicable.

## Institutional Review Board Statement

Not applicable.

## Informed Consent Statement

Not applicable.

## Data Availability Statement

The original contributions presented in the study are included in the article/supplementary material, further inquiries can be directed to the corresponding author.

## Conflict of Interest

The authors declare that the research was conducted in the absence of any commercial or financial relationships that could be construed as a potential conflict of interest.

## Publisher's Note

All claims expressed in this article are solely those of the authors and do not necessarily represent those of their affiliated organizations, or those of the publisher, the editors and the reviewers. Any product that may be evaluated in this article, or claim that may be made by its manufacturer, is not guaranteed or endorsed by the publisher.

## Author Contributions

Not applicable.

## About the Authors

**LI, Keqin**

AMA University, Philippines.

**ZHU, Armando**

Carnegie Mellon University, USA.

**ZHAO, Peng**

Affiliation: Microsoft, China.

**SONG, Jintong**

Computer Science, Boston University, Boston, Massachusetts, USA

**LIU, Jiabei**

Northeastern University, USA.